\documentclass[aps, reprint,showpacs,showkeys,pre,nofootinbib]{revtex4-1}
\usepackage{dcolumn}
\usepackage[utf8]{inputenc}
\usepackage{amsmath}
\usepackage{amsfonts}
\usepackage{amssymb}
\usepackage{graphicx}
\usepackage{hyperref}
\usepackage{xcolor}

\newcommand{\bs}{\boldsymbol}
\newcommand{\mb}{\mathbf}

\begin{document}

\title{Hidden momentum in continuous media and the Abraham-Minkowski debate}

\author{Raul Corr\^ea}%
\author{Pablo L. Saldanha}\email{saldanha@fisica.ufmg.br}%

\affiliation{Departamento de F\'isica, Universidade Federal de Minas Gerais, Belo Horizonte, Minas Gerais 31270-901, Brazil}

\date{7 December 2020}

\begin{abstract}
We perform a detailed study of the connection between hidden momentum and the Abraham-Minkowski debate about the electromagnetic momentum density in material media. The results of a previous work on the subject [\href{https://link.aps.org/doi/10.1103/PhysRevA.102.063510}{P. L. Saldanha and J. S. Oliveira Filho}, Phys. Rev. A \textbf{95} 043804 (2017)] are extended to the continuous medium limit, where some subtleties arise. We consider a polarized and magnetized continuous medium with applied electric and magnetic static fields, where the medium polarization can be due to either an electric charge density or a hypothetical magnetic current density and the medium magnetization can be due to either an electric current density or a hypothetical magnetic charge density. Each model leads to a different expression for the system material hidden momentum and for the electromagnetic momentum density in the medium. We show that the main results of the cited reference are sustained in the continuous medium limit: Abraham momentum is compatible with a model for the medium where the polarization is due to electric charges and the magnetization is due to magnetic charges, Minkowski momentum is compatible to a model where the polarization is due to magnetic currents and the magnetization is due to electric currents, and the expression $\varepsilon_0\mb{E}\times\mb{B}$ is compatible with a model where the polarization is due to  electric charges and the magnetization is due to electric currents, which is the natural model. These results are illustrated with the example of a uniformly polarized and magnetized sphere.
\end{abstract}

\maketitle

\section{Introduction}

The momentum of light in a material medium has been the subject of extensive discussion for more than a century, in the so-called Abraham-Minkowski debate \cite{brevik79,pfeifer07,barnett10b,milonni10,kemp11,mcdonald17}. Early in the preceding century, Abraham proposed that the momentum density of an electromagnetic wave in a medium should be $\mb{E}\times\mb{H}/c^2$, while Minkowski proposed the expression $\mb{D}\times\mb{B}$. In these expressions, $\mb{E}$ represents the electric field, $\mb{B}$ is the magnetic field, $\mb{D}\equiv\varepsilon_0\mb{E}+\mb{P}$, $\mb{H}\equiv\mb{B}/\mu_0 - \mb{M}$, $\mb{P}$ and $\mb{M}$ are the polarization and magnetization of the medium, respectively, $\varepsilon_0$ and $\mu_0$ are the permittivity and permeability of free space, respectively, and $c$ is the speed of light in vacuum. A possible solution for the dilemma, with which we agree, is that there are different ways to divide the total energy-momentum tensor of an electromagnetic wave in a medium into electromagnetic and material tensors such that Abraham's and Minkowski's expressions for the electromagnetic momentum density are related to different divisions of the total energy-momentum tensor \cite{pfeifer07,penfield}. More recently, Abraham's expression was associated with the kinetic electromagnetic  momentum density and Minkowski's with the canonical electromagnetic momentum density \cite{barnett10a,saldanha17}. Another valuable expression for the electromagnetic part of the momentum density of an electromagnetic wave in a medium is $\varepsilon_0\mb{E}\times\mb{B}$, which was shown to be compatible with momentum conservation in several situations when the material part of the momentum is computed by means of the Lorentz force law \cite{saldanha10,saldanha11}.  

In a recent work, a connection between the Abraham-Minkowski debate and the concept of hidden mechanical momentum, regarding the models used for the medium polarization and magnetization, was established \cite{saldanha17}. The relation between hidden momentum and the Abraham-Minkowski debate from other perspectives was considered before \cite{penfield,saldanha10,saldanha11,mcdonald12,saldanha13}. Hidden momentum is a relativistic effect that causes a magnetic dipole $\mb{m}$ in the presence of an external electric field $\mb{E}$ to have a linear momentum $\mb{m}\times\mb{E}/c^2$ even if it is not moving \cite{babson09}. Several classical \cite{penfield,shockley67,vaidman90,hnizdo97,hnizdo97b,babson09,mcdonald12b} and quantum \cite{juvenil15,cameron18} models justify its existence. It appears due to the influence of the external electric field on the current density that composes the magnetic dipole and is thus associated with internal moving parts of the dipole. It is important to mention that in a stationary system whose center of energy is at rest, the hidden mechanical momentum is exactly compensated by the electromagnetic momentum such that the system total momentum is zero, as it must be \cite{babson09}.

In Ref. \cite{saldanha17}, different models for the electric and magnetic dipoles that compose a material medium in the presence of an electromagnetic field were considered. The electric dipoles could be formed either by electric charges or by hypothetical magnetic currents, and the magnetic dipoles could be formed either by electric currents or by hypothetical magnetic charges. Depending on the model, different expressions for the system material hidden momentum are obtained and different material momenta are associated with different expressions for the electromagnetic part of the momentum density in a medium \cite{saldanha17}. However, the treatment of Ref. \cite{saldanha17} considered individual particles with electric and magnetic dipole moments, disregarding the dependence of the electric and magnetic fields inside the medium on the models for the particles electric and magnetic dipole moments, thus not achieving the continuous medium limit. So that treatment was incomplete and it is not clear if its conclusions are sustained with a continuous medium. Since the Abraham-Minkowski debate deals with the electromagnetic momentum density in continuous media, it is very important to demonstrate that the alleged relation between hidden mechanical momentum and this debate is sustained in the continuous medium limit.

Here we study this problem in much more detail, considering the subtleties that were not taken into account in Ref. \cite{saldanha17} and properly examining the continuous medium limit. We confirm the principal conclusions achieved in Ref. \cite{saldanha17}: Abraham's expression for the electromagnetic momentum density is compatible with a polarization caused by electric charges and a magnetization caused by magnetic charges, Minkowski's  expression is compatible with a polarization caused by magnetic currents and a magnetization caused by electric currents, and the expression $\varepsilon_0\mb{E}\times\mb{B}$ is compatible with a polarization caused by electric charges and a magnetization caused by electric currents, which are the most natural models. However, the expression we obtain here for the electromagnetic momentum density in a medium with a polarization caused by magnetic currents and a magnetization caused by magnetic charges is different from the one obtained in Ref. \cite{saldanha17}, which must be corrected for a continuous medium. We use Romer's example \cite{romer95} of a uniformly magnetized and polarized sphere to illustrate the results. It is valid to stress that, since we only consider static fields in this work, the momentum transferred to the atoms in a medium by the Abraham force and other optical forces that are null for static fields is not present in the models to be treated. 

The paper is organized as follows. In Secs. \ref{sec:AB} and \ref{sec:dipoles} we briefly review the Abraham-Minkowski debate and the hidden mechanical momentum concept, respectively.  In Sec. \ref{sec:polarization} we consider a  polarized and magnetized medium under external electric and magnetic fields and compute the medium  material hidden momentum for the different models for its polarization and magnetization. Based on that, we associate a different expression for the electromagnetic momentum density with each model by stipulating that the total system momentum, composed of its electromagnetic and material (hidden) parts, must be zero. In Sec. \ref{sec:discussion} we discuss our main results and conclude the paper.

\section{Abraham-Minkowski debate}\label{sec:AB}

Minkowski's formulation of electrodynamics in continuous media predicts an electromagnetic momentum density $\mb{D}\times\mb{B}$ inside a medium \cite{brevik79,pfeifer07}. This expression for the momentum density is associated with an electromagnetic wave momentum proportional to the refractive index $n$ of the medium. Abraham presented another formulation for electrodynamics in continuous media with an electromagnetic momentum density $\mb{E}\times\mb{H}/c^2$, associated with an electromagnetic wave momentum inversely proportional to $n$ \cite{brevik79,pfeifer07}. This was the origin of the so-called Abraham-Minkowski debate, which has lasted for more than a century now \cite{brevik79,pfeifer07,barnett10b,milonni10,kemp11,mcdonald17}. 

At first sight it seems that a simple experiment could solve the dilemma, by testing if light momentum is proportional to $n$ or to $1/n$ in a medium. Jones and Leslie \cite{jones51,jones78} performed experiments measuring the radiation pressure of light on mirrors immersed in liquids, showing that it is proportional to $n$ as predicted by Minkowski's formulation. However, it is also possible to deduce this behavior by using the Abraham form for the electromagnetic momentum and computing a material momentum that accompanies the wave by means of the Lorentz force on the charges of the medium \cite{gordon73,loudon02}. Gibson \textit{et al.} \cite{gibson80} measured the radiation
pressure on the charges of a semiconductor via the photon drag effect, finding a value proportional to $n$, but this result can also be obtained with the use of Abraham momentum \cite{loudon05}. Campbell \textit{et al.} \cite{campbell05} measured the recoil momentum of atoms in a gas after absorbing one photon, also finding a momentum proportional to $n$, but this result can also be obtained with the use of Abraham momentum \cite{leonhardt06}. Ashkin and Dziedzic \cite{ashkin73} measured the radiation pressure on the free surface of a liquid dielectric. With the incidence of a laser beam from air to water, the idea was to see if the surface is pulled up, indicating that the momentum of light increases by entering the medium as predicted by Minkowski's expression, or if the surface is pushed down,  indicating that the momentum of light decreases by entering the medium as predicted by Abraham's expression. The experiments showed a pulled surface \cite{ashkin73}, but with a detailed analysis taking into account the behavior of the radiation forces in the liquid and of its pressure, it is possible to justify this result with the Abraham expression for the momentum density \cite{gordon73}. The experiments and simulations by Astrath \textit{et al.} \cite{astrath14} showed the time behavior of the liquid surface in this situation. In addition, the experiments by Zhang \textit{et al.} \cite{zhang15} showed how, depending on the fluid dynamics, different momentum transfers between light and the fluid can be observed. As pointed out by Brevik \cite{brevik18}, these experiments with surface forces cannot discriminate between Abraham's and Minkowski's formulations, since they involve force terms that are present in both formulations. On the other hand, the experiments by Walker \textit{et al.} \cite{walker75b,walker77} that measured the torque on a dielectric disk suspended in a torsion fiber subjected to external magnetic and electric fields are consistent with the Abraham force present in Abraham's formulation, but these results can also be deduced with Minkowski's formulation \cite{israel77}.

A possible solution for the Abraham-Minkowski debate, with which we agree, is that there are different ways to divide the total energy-momentum tensor of an electromagnetic wave in a medium into electromagnetic and material tensors, such that Abraham's and Minkowski's expressions for the electromagnetic momentum density are related to different divisions of the total energy-momentum tensor \cite{pfeifer07,penfield}. When the proper material tensor is taken into account, the formulations from Abraham, Minkowski, and others predict the same experimental results. The explanations for the experiments described in the preceding paragraph with both formulations support this interpretation. In this way, there can be different expressions for the momentum density, which are associated with different expressions for the force density and consequently with different models for the material medium, in consistent ways. 

However, this is not the end of the debate. The understanding of what the momentum transfers from light to a medium are in different circumstances is important for many applications \cite{kemp11}. In addition, the understanding of the physical meaning of the different expressions for the momentum density is also relevant from a fundamental point of view. Barnett's association of the Abraham momentum with the kinetic momentum of light and of the Minkowski momentum with the canonical momentum of light is a very interesting perspective \cite{barnett10a} (for a different argumentation, see Ref. \cite{saldanha17}). Reinforcing this view, we have previously shown that a quantum treatment of the reflection of a photon by a mirror with quantized motion leads naturally to the Minkowski momentum, due to canonical features of quantum theory \cite{correa16}. Also, an analysis of kinetic versus canonical quantities related to the angular momentum of light in dispersive media, in association with the Abraham-Minkowski debate, has been made by Bliokh \textit{et al.} \cite{bliokh17a,bliokh17b}. Kemp and Sheppard have shown subtleties that appear in the treatment of dispersive negative index metamaterials \cite{kemp17}. The works by Partanen \textit{et al.} \cite{partanen17a,partanen17b,partanen18,partanen19a,partanen19b} that treat the light propagation in a medium using a mass-polariton quasiparticle approach also present important advances. The covariance of the obtained theory is sustained by numerical calculations showing the appearance of a mass density wave propagating with the light pulse. The momentum of this mass density wave corresponds to the difference between Minkowski and Abraham momenta \cite{partanen17a}, providing a very nice interpretation for these two momenta as the total momentum and the electromagnetic momentum of a light pulse in the medium, respectively. Previous works by one of us \cite{saldanha10,saldanha11} have also shown that if the Lorentz force law is used to compute the momentum transfer from light to the medium, the electromagnetic momentum is given by $\varepsilon_0\mb{E}\times\mb{B}$, which agrees with Abraham momentum in nonmagnetic media, but differs when there is magnetization. So the Abraham-Minkowski debate still persists. In this work we discuss how the models for the electric and magnetic dipoles that compose a material medium are associated with different expressions for the material momentum densities and with the electromagnetic momentum densities in the medium, contributing to the understanding of the role of different electromagnetic momentum expressions in the Abraham-Minkowski debate.

\section{Hidden momentum}\label{sec:dipoles}

Let us start this section by presenting Penfield and Haus's model to discuss hidden momentum \cite{penfield} (see also Ref. \cite{saldanha11}). Consider a rectangular loop of sides $a$ and $b$, which carries a current $I$ and is subject to a uniform external electric field $\mb{E}=E\mb{\hat{x}}$, as depicted in Fig. 1. Due to the closed current, the loop has a magnetic dipole $\mb{m} = abI \mb{\hat{y}}$. When the positive charges in the loop go upward on the right side of the rectangle, the electric field does positive work on them; when they go downward on the left side the work is negative (an inversion of the sign of the charges does not alter the final conclusions). Thus a positive charge $q$ moving to the left in the top section of the loop has an energy $qEb$ higher than when it moves to the right in the bottom section. This causes in the loop an energy flux towards the left. If there is a spatial density of $N$ such loops per unit volume, by noting that the energy per unit time flowing over a length $a$ towards the left for each loop is $IEb$, we conclude that the energy flux in this medium is
\begin{align}
\mb{S}_{\text{mat}} = -IEbaN\mb{\hat{z}} = \mb{M}\times\mb{E},\label{s-loop}
\end{align}
where $\mb{M} = abIN\mb{\hat{y}}$ is the medium magnetization. In this way, in a general system we can divide the system energy flux $\mb{E}\times\mb{H}$ into the mechanical part $\mb{S}_{\text{mat}}$ above and a remaining electromagnetic part $\mb{E}\times\mb{B}/\mu_0$. 

\begin{figure} \begin{center}
  \includegraphics[width=8.5cm]{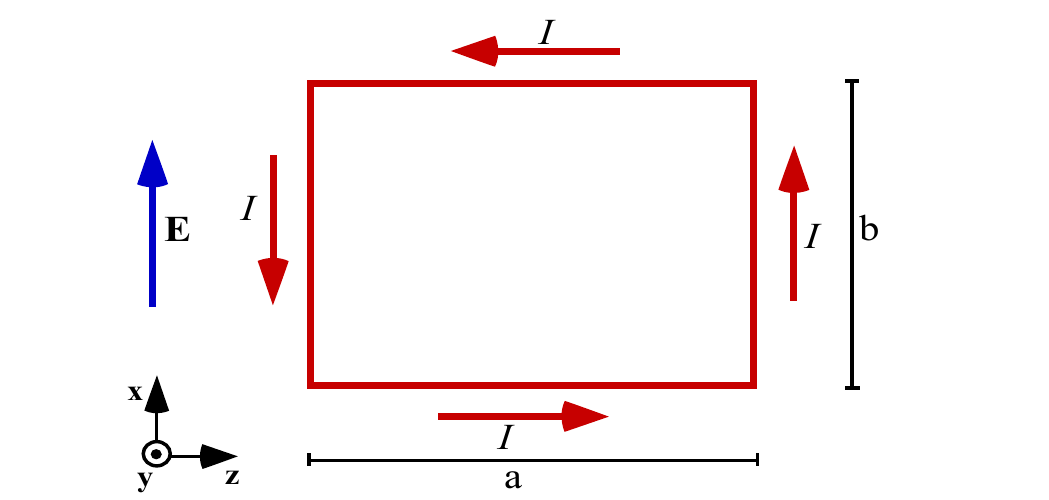}\\
  \caption{A magnetic dipole formed by electric currents submitted to an external electric field has associated hidden energy flux and hidden linear momentum.}\label{fig-hidden}
\end{center}
 \end{figure}

When it comes to momentum, the higher energy of the particles in the upper section of the rectangle in Fig. \ref{fig-hidden} means that they also have a higher momentum than the bottom ones. In addition, since the ratio between the momentum and the energy of a relativistic particle is $\mb{v}/c^2$, where $\mb{v}$ is the particle velocity, we can calculate the momentum associated with the energy flux $\mb{S}_{\text{mat}}$. We consider that the current $I$ is uniform throughout the loop, so if at the top of the loop a linear charge density $\lambda_l$ flows towards the left with velocity $v_l$, while a charge density $\lambda_r$ at the bottom flows towards the right with velocity $v_r$, we get $I=\lambda_lv_l=\lambda_rv_r$. Denoting by $U$ the energy of a particle traveling at the bottom of the loop and by $q$ the total charge of the circuit, the total relativistic linear momentum of the loop is
\begin{align}
\mb{P} = \left[\frac{\lambda_rv_raU}{qc^2}-\frac{\lambda_lv_la(U+qEb)}{qc^2}\right]\mb{\hat{z}}=-\frac{IabE}{c^2}\mb{\hat{z}}=\frac{\mb{m}\times\mb{E}}{c^2}.
\end{align}
Since the center of mass of the loop is still, the above is called the hidden momentum associated with the interaction between the magnetic dipole and the external field and we see that it is a relativistic effect due to internal moving parts of the circuit. For the medium with $N$ such loops per unit volume, the relativistic momentum density, which is the hidden momentum density, is ${\mb{M}\times\mb{E}/c^2}$.

Although the presented model for hidden momentum may seem too simple, different classical \cite{penfield,shockley67,vaidman90,hnizdo97,hnizdo97b,babson09,mcdonald12b} and quantum \cite{juvenil15,cameron18} models predict this behavior. The system material hidden momentum is exactly compensated by the electromagnetic momentum, such that the system total momentum is zero, as it should be in a stationary system whose center-of-energy is at rest. We discuss this issue in the following.

For static electric and magnetic fields, the system electromagnetic momentum can be written as \cite{griffiths18}
\begin{eqnarray}\label{pemee}\nonumber
	\mb{P}_{EM}&=&\int \varepsilon_0\mb{E}\times\mb{B}\,dV=-\int \varepsilon_0(\boldsymbol{\nabla}\Phi)\times\mb{B}\,dV \\\nonumber
	           &=& -\int \varepsilon_0\boldsymbol{\nabla}\times (\Phi\mb{B})\,dV+ \int \varepsilon_0\Phi (\boldsymbol{\nabla}\times \mb{B})\,dV \\ &=&-\mb{P}_{hid},
\end{eqnarray}
where
\begin{equation}\label{hid2}
	\mb{P}_{hid}=-\frac{1}{c^2}\int \Phi \mb{J}dV
\end{equation}
is the system hidden mechanical momentum, with $\mb{J}$ the electric current density and $\Phi$ the electrostatic potential. Note that in the deduction of Eqs. (\ref{pemee}) and (\ref{hid2}) there was no restriction on the medium properties. So these equations are valid in any medium, in particular in dielectric and magnetic media, as long as $\mb{J}$ contains the free as well as the bound currents in the media. Note also that the medium does not need to be uniform. In the following sections we will consider the fields of a uniformly polarized and magnetized sphere as an example. In this case the electromagnetic momentum of the first line of Eq. (\ref{pemee}) must be computed by integrating the momentum density both inside and outside the sphere. 

If the current density is due to the medium magnetization, with $\mb{J}=\bs{\nabla}\times\mb{M}$, by using the vector identity $\bs{\nabla}\times(\Phi\mb{M})=\Phi(\bs{\nabla}\times\mb{M})-\mb{M}\times(\bs{\nabla}\Phi)$, Eq. (\ref{hid2}) can be written as
\begin{equation}\label{hid3}
	\mb{P}_{hid}=\int \frac{\mb{M}\times\mb{E}}{c^2}\,dV
\end{equation}
such that we have a material hidden momentum density $\mb{M}\times\mb{E}/c^2$ in the medium. For a magnetic dipole $\mb{m}$ at the position $\mb{r}$, Eq. (\ref{hid3}) reduces to $\mb{P}_{hid}={\mb{m}\times\mb{E}(\mb{r})}/{c^2}$. The system hidden mechanical momentum is compensated by the electromagnetic momentum if the expression for the electromagnetic momentum density is  $\varepsilon_0\mb{E}\times\mb{B}$, as in Eq. (\ref{pemee}). Since the total momentum of a stationary system whose center of energy is at rest must be zero, this result is consistent. 

It is valid to stress that the hidden momentum cannot be measured in a direct way. For instance, Mansuripur has presented a paradox regarding the torque exerted by an electric charge on a magnetic dipole, questioning the validity of the Lorentz force law \cite{mansuripur1}. By computing the torque using the Lorentz force law, he found a zero torque in a reference frame where both the charge and the dipole are at rest, but a nonzero torque in other frames. With the use of the Einstein-Laub force, he found a null torque in any frame. However, later it was shown that, by considering the magnetic dipole hidden momentum in the presence of the charge electric field, the torque was equal to the time derivative of the system hidden angular momentum in any frame \cite{saldanha13,vanzella1,khorrami1,barnett1}. So both treatments are consistent: the use of the Lorentz force with the presence of hidden momentum in the magnetic dipole or the  use of the Einstein-Laub force without hidden momentum in the dipole. This example illustrates how the model for a magnetic dipole is connected to the force equation that must be used. In principle, it should be possible to verify the hidden momentum influence in an indirect way, in a situation like the one used by Shockley and James to introduce the hidden momentum concept \cite{shockley67}. By varying the system electric field or magnetization in some situations, the system initial hidden momentum could be converted into a kinetic momentum. However, the magnitude of this kinetic momentum would be very small such that this effect was never observed. In principle, it is possible that the same final kinetic momentum may be predicted with the use of another force equation and disregarding the system hidden momentum such that this indirect observation would not prove that the system had hidden momentum before. Despite its nonmeasurability, hidden momentum is an important concept for the electromagnetic theory, being necessary for the validity of the Lorentz force law, for instance.

If, on the other hand, the magnetic dipole is formed not by electric currents but by hypothetical magnetic charges, the system cannot have hidden mechanical momentum since no work is done by the electric field on the static magnetic charges that compose the magnetic dipole. For a similar reason an electric dipole formed by electric charges in the presence of an external magnetic field also does not have hidden momentum. In contrast, an electric dipole $\mb{p}$ formed by hypothetical magnetic currents in the presence of an external magnetic field $\mb{B}$ would have a hidden mechanical momentum given by $-\mb{p}\times \mb{B}$ \cite{griffiths18,griffiths92}. The origin of this hidden momentum is that the magnetic field would do different amounts of work on the magnetic currents that compose the dipole at different parts of the circuit, similarly to the hidden momentum of ordinary magnetic dipoles in the presence of an electric field.

\section{Hidden momentum and the electromagnetic momentum density}\label{sec:polarization} 

We now discuss the relation between the model used for a polarized and magnetized medium, the system hidden mechanical momentum, and the expression for the electromagnetic momentum density, when the hypothetical existence of magnetic charges and currents is taken into account. We consider the case of a polarized and magnetized material medium localized in a finite region of space, at rest. A known external static electric field $\mb{E}_0$ is present due to a known charge density $\rho_0$ outside this region and a known external static  magnetic field $\mb{B}_0$ is present due to a known current density $\mb{J}_0$ outside this region. The medium polarization $\mb{P}$ and magnetization $\mb{M}$ are also known and could have been induced by the external fields or not. The system total electric field is the superposition of $\mb{E}_0$ with the field generated by the medium polarization and the system total magnetic field is the superposition of $\mb{B}_0$ with the field generated by the medium magnetization. We compute the system hidden momentum for different models used for the medium polarization and magnetization. The polarization can be caused either by electric charges or by hypothetical magnetic currents, while the magnetization can be caused either by electric currents or by hypothetical magnetic charges.  

In the region outside the medium, the electric and magnetic fields are independent of the models adopted for the medium polarization and magnetization such that the electromagnetic momentum outside the medium is independent of the model. In general, there will be hidden momentum in the current density $\mb{J}_0$ that creates the external magnetic field $\mb{B}_0$ in the presence of an electric field, and this hidden momentum is also independent of the medium model. On the other hand, the electromagnetic momentum and the system hidden momentum inside the medium do depend on the model for the system polarization and magnetization. The idea of our treatment is to impose that the sum of the hidden momentum density and the electromagnetic momentum density inside the medium is always the same. In this way, we guarantee that the system total hidden momentum always cancel the system total electromagnetic momentum such that the system total momentum is zero, as it must be, since its center of energy is at rest. Thus each model for the medium polarization and magnetization is associated with a different material hidden momentum density and consequently with a different electromagnetic momentum density. To illustrate the treatment, we apply the results to the case of a uniformly polarized and magnetized sphere in the absence of external fields. 

\subsection{Model: Polarization due to electric charges and magnetization due to electric currents}\label{ssee}

First we will consider the model in which the polarization is caused by electric charges and the magnetization is caused by electric currents in the medium, which is the natural one. In this case we have a bound electric charge density $-\bs{\nabla}\cdot\mb{P}$ and a bound electric current density $\bs{\nabla}\times\mb{M}$ in the medium. We can then write Maxwell's equations for the static regime without free charges and currents in the usual manner \cite{griffiths}
\begin{align}
\boldsymbol{\nabla}\cdot(\varepsilon_0\mb{E}&+\mb{P}) = 0, \ \ \ \ \ -\boldsymbol{\nabla}\times(\mb{E}/\mu_0) = 0,\notag\\
\boldsymbol{\nabla}\cdot(\mb{B}&/\mu_0) = 0, \ \ \ \ \ \ \boldsymbol{\nabla}\times(\mb{B}/\mu_0 - \mb{M}) = 0.\label{maxee}
\end{align}
By knowing the external charge and current densities $\rho_0$ and $\mb{J}_0$, as well as the medium polarization and magnetization $\mb{P}$ and $\mb{M}$, as we assume in the present work, we can compute the electric and magnetic fields that are the solutions of the above equations inside the medium, calling these fields $\mb{E}_e$ and $\mb{B}_e$, respectively. We have put a subscript $e$ on both fields to indicate that they correspond to the superposition of the external fields $\mb{E}_0$ and $\mb{B}_0$ with the fields produced by bound \emph{electric} charges and currents in the medium.

This case was treated in the preceding section. The system hidden momentum is given by Eq. (\ref{hid2}), being associated with a hidden momentum density
\begin{equation}\label{hideed}
	{\mb{p}}_{hid}^{(ee)}=\frac{\mb{M}\times\mb{E}_e}{c^2}
\end{equation}
in the medium. The superscript $(ee)$ indicates that the medium polarization is caused by \textit{electric} charges (first index) and the medium magnetization is caused by \textit{electric} currents (second index). 

The material hidden momentum is compensated by the electromagnetic momentum given by Eq. (\ref{pemee}). So the expression
\begin{equation}\label{EMeed}
	{\mb{p}}^{(ee)}_{EM}=\varepsilon_0\mb{E}_e\times\mb{B}_e
\end{equation}
for the electromagnetic momentum density in a medium is compatible with the natural models for the medium polarization and magnetization. In the above expression, the superscript $(ee)$ indicates that the electric field is a combination of the external field with a field caused by \textit{electric} charges in the medium (first index) and the magnetic field is a combination of the external field with a field caused by \textit{electric} currents in the medium (second index).

In the Abraham-Minkowski debate, the electric and magnetic fields inside a medium are usually computed with the traditional form of Maxwell's equations, with no magnetic charges or currents, which reduce to Eq. (\ref{maxee}) in a medium with no free charges or currents in the static regime. So these fields are computed as $\mb{E}_e$ and $\mb{B}_e$ in the situation considered in this work, where the external charge and current densities and the medium polarization and magnetization are supposed to be known. In the same way, the electric displacement $\mb{D}$ and the field $\mb{H}$ are computed as $\mb{D}_e= \varepsilon_0\mb{E}_e + \mb{P}$ and $\mb{H}_e= {\mb{B}_e}/{\mu_0}- \mb{M}$, respectively. For this reason,  in the rest of our paper we will use the expression $\mb{E}_e\times\mb{H}_e/c^2$ as the Abraham electromagnetic momentum and the expression $\mb{D}_e\times\mb{B}_e$ as the Minkowski electromagnetic momentum.

To illustrate the results presented in this section, let us apply them to the case of a uniformly polarized and magnetized sphere in the absence of external fields ($\mb{E}_0=0$ and $\mb{B}_0=0$) \cite{romer95}. Consider a sphere with radius $R$, uniform polarization $\mb{P}$, and uniform magnetization $\mb{M}$. Outside the sphere, the electric and magnetic fields can be computed independently of the model for the polarization and magnetization of the sphere, where \cite{griffiths}
\begin{equation}\label{Eext}
	\mb{E}_{ext}=\frac{1}{4\pi\varepsilon_0 r^3}\left[3(\mb{p}\cdot\mb{\hat{r}})\mb{\hat{r}}-\mb{p}\right],
\end{equation}
\begin{equation}\label{Bext}
	\mb{B}_{ext}=\frac{\mu_0}{4\pi r^3}\left[3(\mb{m}\cdot\mb{\hat{r}})\mb{\hat{r}}-\mb{m}\right],
\end{equation}
with $\mb{p}$ and $\mb{m}$ the electric and magnetic dipole moments of the sphere, which are related to the sphere polarization and magnetization through the relations
\begin{equation}\label{P}
	\mb{P}=\frac{3\mb{p}}{4\pi R^3},
\end{equation}
\begin{equation}\label{M}
	\mb{M}=\frac{3\mb{m}}{4\pi R^3}.
\end{equation}
The momentum carried by the electromagnetic field outside the sphere is given by \cite{griffiths18}
\begin{equation}\label{pext}
	\mb{P}_{ext}=\int_{r>R}\varepsilon_0\mb{E}_{ext}\times\mb{B}_{ext} \;dV = -\frac{\mu_0}{4\pi R^3}\frac{\mb{p}\times{\mb{m}}}{3}.
\end{equation}

If the sphere polarization is caused by a surface electric bound charge  density (the usual), the electric field inside the sphere is uniform, given by \cite{griffiths}
\begin{equation}\label{Ee}
	\mb{E}_e=-\frac{\mb{p}}{4\pi\varepsilon_0R^3}.
\end{equation}
The electric displacement is given by
\begin{equation}\label{De}
	\mb{D}_e= \varepsilon_0\mb{E}_e + \mb{P} = \frac{2\mb{p}}{4\pi R^3},
\end{equation}
where Eq. (\ref{P}) was used.

If the sphere magnetization is caused by a surface electric bound current density (as is usual), the magnetic field inside the sphere is also uniform, given by \cite{griffiths}
\begin{equation}\label{Be}
	\mb{B}_e=\frac{2\mu_0\mb{m}}{4\pi R^3}.
	\end{equation}
The vector $\mb{H}$ is given by
\begin{equation}\label{He}
	\mb{H}_e= \frac{\mb{B}_e}{\mu_0}- \mb{M} = -\frac{\mb{m}}{4\pi R^3},
\end{equation}
where Eq. (\ref{M}) was used.

In the present model, each magnetic dipole $\mb{m}_i$ that composes the medium is subjected to a uniform electric field and carries a hidden momentum $\mb{m}_i\times\mb{E}_e/c^2$. This fact results in a hidden momentum density in the medium given by Eq. (\ref{hideed}) such that the system hidden mechanical momentum can be computed as 
\begin{equation}\label{hidee}
	\mb{P}_{hid}^{(ee)}=\int\frac{\mb{M}\times\mb{E}_e}{c^2}dV=\frac{\mu_0}{4\pi R^3}\mb{p}\times\mb{m},
\end{equation}
where Eqs. (\ref{M}) and (\ref{Ee}) were used.  Considering the electromagnetic momentum density from Eq. (\ref{EMeed}), by using Eqs. (\ref{Ee}), (\ref{Be}), and (\ref{pext}), the electromagnetic momentum is
\begin{equation}\label{EMee}
	\mb{P}_{EM}^{(ee)}=\int_{r<R}\varepsilon_0\mb{E}_e\times\mb{B}_edV+ 	\mb{P}_{ext} =-\frac{\mu_0}{4\pi R^3}\mb{p}\times\mb{m}
\end{equation}
such that the electromagnetic momentum cancels the material momentum of Eq. (\ref{hidee}). This example illustrates the fact that the hidden momentum of a stationary system whose center of energy is at rest is exactly compensated by the electromagnetic momentum, as expressed in Eq. (\ref{pemee}).

Another way to compute the system hidden mechanical momentum in this case is to use Eq. (\ref{hid2}), with $\Phi$ the electrostatic potential generated by the polarized medium and $\mb{J}$ the bound current associated with the magnetized medium. For a uniformly magnetized sphere, the bound current corresponds to a surface current $\mb{K}=\mb{M}\times\mb{\hat{n}}$ at the sphere surface, where $\mb{\hat{n}}$ is the normal unit vector \cite{griffiths}. The electrostatic potential at the surface of a uniformly polarized sphere with dipole moment in the $\mb{\hat{z}}$ direction can be written as $\Phi=(\mb{P}\cdot\mb{\hat{n}})R/(3\varepsilon_0)$ \cite{griffiths}. So the system hidden momentum can be computed as
\begin{eqnarray}\label{hideeb}\nonumber
	\mb{P}_{hid}^{(ee)}&=&-\frac{R^3}{3\varepsilon_0c^2}\int_0^{2\pi}d\phi\int_0^\pi \sin(\theta) d\theta\,(\mb{P}\cdot\mb{\hat{n}}) \,(\mb{M}\times\mb{\hat{n}})\\
	            &=&\frac{\mu_0}{4\pi R^3}\mb{p}\times\mb{m},
\end{eqnarray}
resulting in the same value of Eq. (\ref{hidee}).

\subsection{Model: Polarization due to electric charges and magnetization due to magnetic charges}\label{ssem}

We consider now that the medium polarization is still due to electric charges, while its magnetization is due not to electric currents but to hypothetical magnetic charges. In this case we have a bound electric charge density ${-\bs{\nabla}\cdot\mb{P}}$ and a bound magnetic charge density $-\bs{\nabla}\cdot\mb{M}$ in the medium \cite{mcdonald2}.  With such a model, Maxwell's equations for the static regime without free charges and currents are \cite{mcdonald2}
\begin{align}
\boldsymbol{\nabla}\cdot(\varepsilon_0\mb{E}+&\mb{P}) = 0, \ \ \ \ \ -\boldsymbol{\nabla}\times(\mb{E}/\mu_0) = 0,\notag\\
\boldsymbol{\nabla}\cdot(\mb{B}/\mu_0+&\mb{M}) = 0, \ \ \ \ \ \ \ \boldsymbol{\nabla}\times(\mb{B}/\mu_0) = 0.\label{maxem}
\end{align}
If we compare Maxwell's equations for both models presented so far, we see that the replacement of $\mb{B}/\mu_0 - \mb{M}$ by $\mb{B}/\mu_0$ takes us from Eqs. \eqref{maxee} to Eqs. \eqref{maxem}. We recall that we consider the same polarization $\mb{P}$ and magnetization $\mb{M}$ as in Sec. \ref{ssee}, though the magnetization is now caused by a different model. Thus, if we have already solved the system of equations \eqref{maxee}, finding the electric and magnetic fields $\mb{E}_e$ and $\mb{B}_e$ as the solutions, we can compute the magnetic field for the new model as $\mb{B}_m/\mu_0=\mb{B}_e/\mu_0 - \mb{M}=\mb{H}_e$, while we have the same electric field $\mb{E}_e$ as before. The subscript $m$ in the new solution for the magnetic field indicates that it corresponds to the superposition of the external field $\mb{B}_0$ with the field produced by \emph{magnetic} charges in the medium.

It is important to stress that in the present work we consider that the medium magnetization can be caused either only by electric currents, as in Sec. \ref{ssee}, or only by magnetic charges, as in this section. If the magnetization has contributions from both electric currents and magnetic charges, a more complex description must be made, as in Ref. \cite{mcdonald2}. The same happens with the medium polarization, which in this work is considered to be caused either only by electric charges or only by magnetic currents.

In this section's model, there is no hidden momentum density in the medium, as discussed at the end of Sec. \ref{sec:dipoles}, and the total material hidden momentum in the medium is 
\begin{equation}\label{hidem}
	\mb{P}_{hid}^{(em)}=0.
\end{equation}
The superscript $(em)$ indicates that the medium polarization is caused by \textit{electric} charges (first index) and the medium magnetization is caused by \textit{magnetic} charges (second index). 

The different expressions for the electromagnetic part of the momentum density in the Abraham-Minkowski debate are associated with different divisions of the total system energy-momentum tensor into a material tensor and an electromagnetic tensor \cite{pfeifer07,penfield}. In addition, the form of the material tensor is intimately connected with the physical model used for the medium \cite{penfield}. In the model of Sec. \ref{ssee} we had a contribution $\mb{M}\times\mb{E}_e/c^2$ for the material momentum density [Eq. (\ref{hideed})]. In the model of the present section, this momentum density does not exist in the material tensor, so it must be included in the electromagnetic tensor for the total momentum density to be the same, in order to guarantee that the system total momentum is zero. Adding this quantity to the electromagnetic momentum density of Eq. (\ref{EMeed}), we obtain
\begin{eqnarray}\label{EMemd}
	{\mb{p}}^{(em)}_{EM}&=&\varepsilon_0\mb{E}_e\times\mb{B}_e+\frac{\mb{M}\times\mb{E}_e}{c^2}=\varepsilon_0\mb{E}_e\times\mb{B}_m,
\end{eqnarray}
where we used $\mb{B}_m/\mu_0=\mb{B}_e/\mu_0 - \mb{M}$. The superscript $(em)$ indicates that the electric field is a combination of the external field with a field caused by \textit{electric} charges in the medium (first index) and the magnetic field is a combination of the external field with a field caused by \textit{magnetic} charges in the medium (second index). With this electromagnetic momentum density, we have
\begin{eqnarray}\label{pememr}\nonumber
	\mb{P}_{EM}^{(em)}&=&\int \varepsilon_0\mb{E}_e\times\mb{B}_m\,dV=-\int \varepsilon_0(\boldsymbol{\nabla}\Phi_e)\times\mb{B}_m\,dV \\\nonumber
	           &=& \int \varepsilon_0\Phi_e (\boldsymbol{\nabla}\times \mb{B}_m)\,dV = 0,
\end{eqnarray}
where Eq. (\ref{maxem}) was used. We can furthermore compare Eq. (\ref{EMemd}) with the solutions of Sec. \ref{ssee} by recalling that $\mb{B}_m/\mu_0=\mb{H}_e$, such that
\begin{eqnarray}\label{EMemd-abraham}
	{\mb{p}}^{(em)}_{EM}&=&\frac{\mb{E}_e\times\mb{H}_e}{c^2},
\end{eqnarray}    
which is Abraham momentum density.

Applying these results to the case of a uniformly polarized and magnetized sphere without external fields, by using Eqs. (\ref{Ee}), (\ref{He}), and (\ref{pext}) we have
\begin{equation}\label{EMem}
	\mb{P}_{EM}^{(em)}=\int_{r<R}\frac{\mb{E}_e\times\mb{H}_e}{c^2}\, dV+ 	\mb{P}_{ext} = 0.
\end{equation}

Note that these models for the medium polarization and magnetization are compatible with the system material momentum having only a kinetic origin, since there is no hidden mechanical momentum in the medium. In this way, the material momentum is nonzero only if the medium moves. As we see, the Abraham momentum for the electromagnetic field is compatible with the material momentum being purely kinetic, so it can be associated with the kinetic electromagnetic momentum, as discussed in previous works \cite{barnett10a,saldanha17}.

It is worth mentioning that a hidden momentum $\mb{M}\times\mb{E}_e/c^2$ is accompanied by a hidden energy flux $\mb{M}\times\mb{E}_e$, as discussed in Sec. \ref{sec:dipoles}. So when we change the model for the material medium removing the contribution of hidden momentum from the momentum density of the material energy-momentum tensor, we also remove the contribution of the hidden momentum flux from the energy flux in the material energy-momentum tensor. In this way, if we have a symmetric and divergence-free total energy-momentum tensor for the material model of Sec. \ref{ssee}, as required by the conservation laws of energy, momentum, and angular momentum, the total energy-momentum tensor for the model of this subsection, as well as for the other models to be treated, is also symmetric and divergence-free.

\subsection{Model: Polarization due to magnetic currents and magnetization due to electric currents}\label{ssme}

In our third model, consider that the polarization is caused by hypothetical magnetic currents and that the magnetization is caused by electric currents. In this case we have a bound electric current density $\bs{\nabla}\times\mb{M}$ and a bound magnetic current density $-c^2\bs{\nabla}\times\mb{P}$ in the medium \cite{mcdonald2}. Maxwell's equations for the static regime without free charges and currents in this model become \cite{mcdonald2}
\begin{align}
\boldsymbol{\nabla}\cdot(\varepsilon_0\mb{E}) = 0, \ \ \ \ \ -&\boldsymbol{\nabla}\times(\mb{E}/\mu_0-c^2\mb{P}) = 0,\notag\\
\boldsymbol{\nabla}\cdot(\mb{B}/\mu_0) = 0, \ \ \ \ \ \ &\boldsymbol{\nabla}\times(\mb{B}/\mu_0-\mb{M}) = 0,\label{maxme}
\end{align}
Comparing to Eqs. (\ref{maxee}), we see that the replacement of $\varepsilon_0\mb{E} + \mb{P}$ by $\varepsilon_0\mb{E}$ takes us from Eqs. \eqref{maxee} to Eqs. \eqref{maxme}. So, recalling that $\mb{P}$ and $\mb{M}$ are the same as in previous sections, if we have already solved the system of equations \eqref{maxee}, finding the electric and magnetic fields $\mb{E}_e$ and $\mb{B}_e$ as solutions, we can compute the electric field in the new model as $\varepsilon_0\mb{E}_m=\varepsilon_0\mb{E}_e + \mb{P}=\mb{D}_e$, while the magnetic field has the value $\mb{B}_e$. The subscript $m$ in the new solution for the electric field indicates that it corresponds to the superposition of the external field $\mb{E}_0$ with the field produced by \emph{magnetic} currents in the medium.

In relation to the model of Sec. \ref{ssee}, we  have an extra hidden momentum density $-\mb{P}\times\mb{B}_e$ in the medium with the present model, since each electric dipole $\mb{p}_i$ that composes the medium is subjected to a uniform magnetic field and carries a hidden mechanical momentum ${-\mb{p}_i\times\mb{B}_e}$ \cite{griffiths18,griffiths92}. Adding this quantity to Eq. (\ref{hideed}), we obtain a hidden momentum density
\begin{equation}\label{hidmed}
	{\mb{p}}_{hid}^{(me)}=\frac{\mb{M}\times\mb{E}_m}{c^2}-{\mb{P}\times\mb{B}_e} +\mu_0\mb{P}\times\mb{M},
\end{equation}
since we have $\mb{E}_e =\mb{E}_m-\mb{P}/\varepsilon_0$. The first term in the above expression can be associated with the hidden momentum of the magnetic dipoles in the presence of the electric field $\mb{E}_m$ and the second term with the hidden momentum of the electric dipoles in the presence of the magnetic field $\mb{B}_e$. The presence of the term $\mu_0\mb{P}\times\mb{M}$ in the above expression is certainly strange, but it is essential for having a zero total momentum in the system, being associated with how the electromagnetic fields are generated by the medium polarization and magnetization and how they contribute to the system hidden momentum. Below we use the example of a uniformly polarized and magnetized sphere to show that the hidden momentum density of Eq. (\ref{hidmed}) results in the correct hidden mechanical momentum for the system. 

Again, we impose that the total momentum density in the medium must be the same for all models, to guarantee that the system total momentum is zero. So in any model $\mb{p}_{EM}+\mb{p}_{hid}$ must result in the same expression. Since we added a contribution $-\mb{P}\times\mb{B}_e$ to the hidden momentum density in relation to the model of Sec. \ref{ssee}, arriving at the expression (\ref{hidmed}), we must subtract this quantity from the electromagnetic momentum density of Eq. (\ref{EMeed}), obtaining 
\begin{eqnarray}\label{EMmed}\nonumber
	\mb{p}^{(me)}_{EM}&=&\varepsilon_0\mb{E}_e\times\mb{B}_e + {\mb{P}\times\mb{B}_e}=\varepsilon_0\mb{E}_m\times\mb{B}_e\\
	                  &=& {\mb{D}_e\times\mb{B}_e},
\end{eqnarray}
with $\varepsilon_0\mb{E}_m=\mb{D}_e$, which is the Minkowski momentum. 

Now we use the case of a uniformly polarized and magnetized sphere to illustrate the expression (\ref{hidmed}) for the hidden momentum density in the model of this section. If the sphere polarization is due to magnetic currents and the sphere magnetization is due to electric currents, we have surface electric and magnetic currents on the sphere due to the uniform magnetization and polarization. Because there are two currents at the surface of the sphere, each one interacting with the field created by the other, there will be a subtlety due to the difference in the fields internal and external to the sphere. A way to clarify this is to consider one current slightly above the other \cite{griffiths18}. Let us consider that the electric current is located slightly above the magnetic current, a situation that is compatible with a uniformly magnetized sphere slightly bigger than the uniformly polarized sphere. We have a contribution $-\mb{P}\times\mb{B}_e$ for the hidden momentum density in the medium, corresponding to the second term of Eq. (\ref{hidmed}), such that the medium polarization gives rise to the following contribution to the system hidden momentum:
\begin{equation}\label{aadf}
	\mb{P}_{hid}^{(me-a)}=-\int{\mb{P}\times\mb{B}_e}dV=-\frac{2\mu_0}{4\pi R^3}\mb{p}\times\mb{m},
\end{equation}
where Eqs. (\ref{P}) and (\ref{Be}) were used. 

The magnetic dipoles, on the other hand, are not subjected to a uniform electric field since there is a discontinuity of the electric field at the surface of the polarized sphere, considered to be slightly smaller than the magnetized sphere. However, the contribution of the magnetization to the system hidden momentum can be computed with the use of Eq. (\ref{hid2}) since the electric field configuration outside the polarized sphere does not depend on the model for the electric polarization and the bound surface electric current is outside the polarized sphere. The association of an electrostatic potential $\Phi$ is artificial in this case, but it generates the same electric field outside the surface as the electrostatic potential vector that should be used \cite{griffiths18}. So the contribution of the magnetization to the system hidden momentum has the same value as in Eq. (\ref{hideeb}):
\begin{equation}
	\mb{P}_{hid}^{(me-b)}=\frac{\mu_0}{4\pi R^3}\mb{p}\times\mb{m}.
\end{equation}
A direct calculation using the electric field configuration, without defining an artificial electrostatic potential, leads to the same result  \cite{griffiths18}. If we want to write the above contribution for the system hidden mechanical momentum in terms of the volume integral of a hidden momentum density, as in Eq. (\ref{hid3}), this hidden momentum density must be ${\mb{M}\times\mb{E}_e}/{c^2}$. This expression comes from the fact that the simplest way to construct the artificial potential $\Phi$ outside the sphere is to impose that it is equal to the real electrostatic potential in the model where the polarization is caused by an electric charge density, with $-\bs{\nabla}\Phi=\mb{E}_e$ everywhere, and we use this relation to go from Eq. (\ref{hid2}) to Eq. (\ref{hid3}). Substituting $\mb{E}_e=\mb{E}_m-\mb{P}/\varepsilon_0$, this hidden momentum density results in the sum of the first and third terms of Eq. (\ref{hidmed}), justifying the presence of the ``strange term'' $\mu_0\mb{P}\times\mb{M}$ in this equation.

The system total material hidden momentum is thus
\begin{equation}\label{hidme}
	\mb{P}_{hid}^{(me)}=\mb{P}_{hid}^{(me-a)}+\mb{P}_{hid}^{(me-b)}=-\frac{\mu_0}{4\pi R^3}\mb{p}\times\mb{m}.
\end{equation}
Eqs. (\ref{hidee}) and (\ref{aadf}) show that the hidden momentum density of Eq. (\ref{hidmed}), with $\mb{E}_m=\mb{E}_e+\mb{P}/\varepsilon_0$,  leads to the above system hidden momentum.

If we consider that the uniformly magnetized sphere is slightly smaller than the uniformly polarized sphere, then the contribution of the electric dipoles to the system hidden momentum will be given by $\mb{P}_{hid}^{(me-b)}$ and the contribution of the magnetic dipoles will be $\mb{P}_{hid}^{(me-a)}$, generating the same total hidden momentum of Eq. (\ref{hidme}). We cannot consider that the magnetized sphere has exactly the same radius as the polarized sphere because the fields are discontinuous at each corresponding surface and the contribution of hidden momentum comes from the surface currents. 

By using Eqs. (\ref{De}), (\ref{Be}), and (\ref{pext}) we have
\begin{equation}\label{EMme}
	\mb{P}_{EM}^{(me)}=\int_{r<R}\mb{D}_e\times\mb{B}_e\, dV+ 	\mb{P}_{ext} =\frac{\mu_0}{4\pi R^3}\mb{p}\times\mb{m},
\end{equation}
such that the integral of the electromagnetic momentum density of Eq. (\ref{EMmed}) over the whole space cancels the material momentum of Eq. (\ref{hidme}).

\subsection{Model: Polarization due to magnetic currents and magnetization due to magnetic charges}\label{ssmm}

In our last model for the medium polarization and magnetization we consider that the polarization is associated with hypothetical magnetic currents and that the magnetization is associated with hypothetical magnetic charges. In this case we have a bound magnetic charge density $-\bs{\nabla}\cdot\mb{M}$ and a bound magnetic current density $-c^2\bs{\nabla}\times\mb{P}$. Maxwell's equations for this model are \cite{mcdonald2}
\begin{align}
\boldsymbol{\nabla}\cdot&(\varepsilon_0\mb{E}) = 0, \ \ \ \ \ -\boldsymbol{\nabla}\times(\mb{E}/\mu_0-c^2\mb{P}) = 0,\notag\\
\boldsymbol{\nabla}\cdot(&\mb{B}/\mu_0+\mb{M}) = 0, \ \ \ \ \ \ \boldsymbol{\nabla}\times(\mb{B}/\mu_0) = 0.\label{maxmm}
\end{align}
Here we must use both ${\mb{B}/\mu_0 - \mb{M}} \rightarrow \mb{B}/\mu_0$ and ${\varepsilon_0\mb{E} + \mb{P}} \rightarrow \varepsilon_0\mb{E}$ in order to go from Eqs. \eqref{maxee} to \eqref{maxmm}, getting the solutions $\mb{B}_m/\mu_0=\mb{B}_e/\mu_0 - \mb{M}=\mb{H}_e$ and $\varepsilon_0\mb{E}_m=\varepsilon_0\mb{E}_e + \mb{P}={\mb{D}_e}$. 

With this model, the hidden momentum density in the medium is given by  
\begin{equation}\label{phidmmd}
	\mb{p}_{hid}^{(mm)}=-\mb{P}\times\mb{B}_m.
\end{equation}
To find the expression for the electromagnetic momentum density in this case we must subtract this quantity from the expression of Eq. (\ref{EMemd})which corresponds to a situation with no hidden momentum,
\begin{eqnarray}\label{EMmmd}\nonumber
	\mb{p}^{(mm)}_{EM}&=&\frac{\mb{E}_e\times\mb{H}_e}{c^2}+\mb{P}\times\mb{B}_m=\varepsilon_0{\mb{E}_m\times\mb{B}_m}\\ 
	&=& \mu_0\mb{D}_e\times\mb{H}_e,
\end{eqnarray}
with $\mb{B}_m/\mu_0=\mb{H}_e$ and $\varepsilon_0\mb{E}_m={\mb{D}_e}$.

Treating the case of a uniformly polarized and magnetized sphere, using Eqs. (\ref{P}) and  (\ref{He}), the total material hidden momentum of the system is 
\begin{equation}\label{phidmm}
	\mb{P}_{hid}^{(mm)}=-\int\mu_0\mb{P}\times\mb{H}_e\,dV=\frac{\mu_0}{4\pi R^3}\mb{p}\times\mb{m}.
\end{equation}
By using Eqs. (\ref{De}), (\ref{He}), and (\ref{pext}) we have
\begin{equation}
	\mb{P}_{EM}^{(mm)}=\int_{r<R}\mu_0\mb{D}_e\times\mb{H}_e\, dV+ 	\mb{P}_{ext} =-\frac{\mu_0}{4\pi R^3}\mb{p}\times\mb{m}    
\end{equation}
such that the electromagnetic momentum cancels the material momentum.

\section{Discussion} \label{sec:discussion}

\begin{table*}\label{tab:table2}
\caption{Association of different models for the medium polarization $\mb{P}$ and magnetization $\mb{M}$ with the corresponding expressions for the hidden mechanical momentum density and the electromagnetic momentum density.}
\begin{ruledtabular}
\begin{tabular}{cccc}
Composition of $\mb{P}$ & Composition of $\mb{M}$ & Hidden momentum density & Electromagnetic momentum density \\
\hline
electric charges & electric currents  &  $\mb{M}\times\mb{E}_e/c^2$ & $\varepsilon_0\mb{E}_e\times\mb{B}_e$ \\
electric charges & magnetic charges  & 0 & $\varepsilon_0\mb{E}_e\times\mb{B}_m=\mb{E}_e\times\mb{H}_e/c^2$ \\
magnetic currents & electric currents  & $\mb{M}\times\mb{E}_m/c^2-\mb{P}\times\mb{B}_e+\mu_0\mb{P}\times\mb{M}$ & $\varepsilon_0\mb{E}_m\times\mb{B}_e=\mb{D}_e\times\mb{B}_e$ \\
magnetic currents & magnetic charges & $-\mb{P}\times\mb{B}_m$ & $\varepsilon_0\mb{E}_m\times\mb{B}_m=\mu_0\mb{D}_e\times\mb{H}_e$ \\\end{tabular}
\end{ruledtabular}
\end{table*}

The main results obtained in the preceding section are summarized in Table I. The electromagnetic momentum densities presented in the first three rows of this table agree with the results presented in Ref. \cite{saldanha17}, but the last row of this table presents a difference in relation to those previous conclusions. As mentioned before, in Ref. \cite{saldanha17} the medium was considered to be composed of nonoverlapping individual particles with electric and magnetic dipole moments and it was not clear if its conclusions persist in the continuous medium limit. In particular, it was not considered that the values of the electric and magnetic fields inside the medium depend on the model for the medium polarization and magnetization and this is the reason for the discrepancy. So the results presented in Table I are the most suitable ones for the electromagnetic momentum densities in a continuous medium. 

Let us compare in more detail the results obtained in Sec. \ref{sec:polarization} for the example of the uniformly polarized and magnetized sphere with the results presented in Ref. \cite{saldanha17}.  The model of Sec. \ref{ssee}, associated with the first row of Table I, is the natural one, without magnetic charges or currents, so the results of both treatments, the one presented in Sec. \ref{sec:polarization} and the one from Ref. \cite{saldanha17}, naturally agree. In the model of Sec. \ref{ssem}, associated with the second row of Table I, there is no hidden momentum in both treatments, independently of the value of the fields, so they also agree. In the model of Sec. \ref{ssme}, associated with the third row of Table I, note that only the electric field inside the polarized sphere changes in relation to the natural model of Sec. \ref{ssee}. So if we consider that the surface of the magnetized sphere is outside the surface of the polarized sphere, the electric current density is submitted to the same external electric field as in the natural model and the magnetic current density is also submitted to the same magnetic field as in the natural model. In this way, one is readily led to write the electromagnetic momentum density for the model of Sec. \ref{ssme} in terms of fields which are solutions of the natural model, so both treatments agree.   It is only in the model of Sec. \ref{ssmm}, associated with the last row of Table I, where both the magnetic field inside the magnetized sphere and the electric field inside the polarized sphere change in relation to the natural model, that this subtlety results in a difference between the treatments. So the expression for the electromagnetic momentum density in a continuous medium with a polarization caused by magnetic currents and a magnetization caused by magnetic charges has the value shown in the last row of Table I. 

Note in Table I  that if we compute the fields $\mb{E}_i$ and $\mb{B}_j$, where $i$ and $j$ can be $e$ or $m$ depending on the models for the medium polarization and magnetization, the electromagnetic momentum density for each model can always be written as $\varepsilon_0\mb{E}_i\times\mb{B}_j$. So the expression $\varepsilon_0\mb{E}\times\mb{B}$ can be considered as the electromagnetic momentum density independently of the model used for the medium polarization and magnetization, as long as we use the corresponding version of Maxwell's equations to compute the fields $\mb{E}$ and $\mb{B}$ in each model. Our results also show that, independently of the model, the $\varepsilon_0\mb{E}\times\mb{B}$ expression leaves the hidden mechanical momentum to the material part of the total momentum density division and this is an important feature of this momentum density expression.

On the other hand, if we compute the fields $\mb{E}$, $\mb{D}$, $\mb{B}$, and $\mb{H}$ inside the medium considering the absence of magnetic charges and magnetic currents in nature, as is usual in the Abraham-Minkowski debate, these fields would be computed as $\mb{E}_e$, $\mb{D}_e$, $\mb{B}_e$, and $\mb{H}_e$, respectively. Then, by checking Table I, we conclude that to guarantee that the system total momentum is zero, by considering that the electromagnetic part of the momentum density is $\varepsilon_0\mb{E}_e\times\mb{B}_e$, the material part of the momentum density must be computed as the hidden mechanical momentum density $\mb{M}\times\mb{E}_e/c^2$, which is compatible with a polarization caused by electric charges and magnetization caused by electric currents. If we consider that the electromagnetic part of the momentum density is $\mb{E}_e\times\mb{H}_e/c^2$, which is Abraham's momentum density, the material part of the momentum must be zero, with no hidden momentum, which is compatible with a polarization caused by electric charges and magnetization caused by magnetic charges. If the electromagnetic part of the momentum density is considered to be $\mb{D}_e\times\mb{B}_e$, which is Minkowski's momentum density, the material part of the momentum must be computed as the hidden momentum $\mb{M}\times\mb{E}_m/c^2-\mb{P}\times\mb{B}_e+\mu_0\mb{P}\times\mb{M}$, which is compatible with a polarization caused by magnetic currents and a magnetization caused by electric currents. These conclusions are the same as the ones from Ref. \cite{saldanha17}.

Considering the absence of magnetic charges and currents in nature, the natural expression for the electromagnetic momentum density is thus $\varepsilon_0\mb{E}\times\mb{B}$. So this expression has its value in the Abraham-Minkowski debate as the one that puts the hidden mechanical momentum in the material momentum part of the total momentum division, besides Abraham's expression being associated with a kinetic electromagnetic momentum and Minkowski's expression with a canonical electromagnetic momentum.

The authors acknowledge David Griffiths, Vladimir Hnizdo, and Kirk McDonald for very useful discussions. This work was supported by the Brazilian agency CNPq.

\end{document}